\documentclass[prl,twocolumn,showpacs,amssymb]{revtex4}
\usepackage{graphicx}
\usepackage{dcolumn}
\usepackage{bm}
\usepackage{color} 
\newcommand{\qvec}{{\bf q}}

\begin{document} 
\title{Time-Dependent Gutzwiller Theory for Multiband Hubbard Models} 
\author{E.~v.~Oelsen, G.~Seibold, and J.~B\"unemann} 
\affiliation{Institut f\"ur Physik, BTU Cottbus, PBox 101344, 03013 Cottbus,  
Germany} 
\date{\today}
 
 
\begin{abstract} 
Based on the variational Gutzwiller theory, we present a method  
for the computation of response 
functions for multiband Hubbard models with general  
local Coulomb interactions.  
The improvement over the conventional  
random-phase approximation is exemplified for an  
infinite-dimensional two-band Hubbard model where 
the incorporation of the local multiplet-structure leads to a much larger  
sensitivity of ferromagnetism on the Hund coupling. 
Our method can be implemented  
into LDA+Gutzwiller schemes and will therefore be an important tool  
for the computation of response functions for strongly correlated materials. 
\end{abstract} 
 

\maketitle 
 
During the last two decades a number of schemes has been developed 
to improve density functional related theories  
(DFT)~\cite{hk64,ks65} for materials with medium to strong 
 Coulomb-interaction effects. For such systems, the standard  
 DFT methods are usually insufficient since they are based on effective  
single-particle approaches.    
In this context, a common idea is to separate the electrons  
in localized orbitals (typically those in d- or f-bands) from  
the delocalized ones and to assume for the former a  
Hubbard-type interaction which is then treated in a 
given approximation.  
In the LDA+U method~\cite{anisimov97} 
the local-density approximation (LDA) 
is combined with a Hartree-Fock (HF) like treatment of the 
local interaction, whereas in the LDA+DMFT approach  
\cite{kotliar06}  
dynamical mean-field theory (DMFT) is applied to deal with 
the Hubbard term. 
The LDA+DMFT method is quite successful in calculating single-particle  
 excitations of correlated materials~\cite{anisimov97}; however, it is  
rather demanding from a numerical 
point of view since it requires the self-consistent solution of  
complex single-impurity models. 
In this regard, the Gutzwiller variational theory is a 
promising alternative to be combined with LDA computations 
 since it joins the numerical simplicity of an LDA+U calculation  
 with the accuracy of DMFT methods.  
The Gutzwiller theory is based on an exact evaluation of Gutzwiller 
 wave functions in the limit of infinite spatial dimensions $D$. 
 This has been accomplished for general multiband Hubbard models 
 in Ref.~\cite{buen98}. Applying the infinite $D$ energy functional to  
finite  dimensional systems is usually denoted as the `Gutzwiller  
approximation' (GA).  
The multi-band GA has been successfully combined with bandstructure  
calculations and, among others, applied to ferromagnetic Nickel  
\cite{buen03,buen08}
and LaOFeAsNa~\cite{schickling11}. 
Self-consistent Gutzwiller-DFT schemes have been developed and applied  
in Refs.\  
\cite{ho08,deng08,wang08,deng09,wang10}. 
 
A comparison of experimental data with results from DFT often requires 
the computation of response functions where 
one is interested in the frequency and wave-vector dependence of 
two-particle correlation functions.
Within DFT, such excitations can be computed via    
the time-dependent DFT (TDDFT), which is based on a proper generalization 
of the Hohenberg-Kohn theorem~\cite{rg84}. In general, when  
the external perturbation is small  
the susceptibility $\chi_q(\omega)$ of the interacting electrons can be 
expressed via the {\it fictitious} response function of the Kohn-Sham  
reference system. 
The structure of the resulting self-consistency equations  
for $\chi_q(\omega)$ bears 
strong resemblance with the random-phase approximation (RPA) where 
the interaction kernel is now composed of the Hartree- and  
exchange-correlation potential. 
 
The difficulties of the DFT to describe materials with strong  
Coulomb interactions equally concern the TDDFT. Therefore, 
 the question arises whether the concept of TDDFT  can be applied to the  
aforementioned improvements of DFT for such systems.  
Concerning DMFT, there are obvious limitations due to the  
 numerical complexity of this approach.  
On the other hand, it has been shown  
that a time-dependent GA (TDGA) can be constructed~\cite{goe01}  
 for single-band Hubbard models. This approach has been successfully applied  
to a number of systems and response functions ~\cite{lor03,goe05,goe08,falk07}. 
In order to construct a time-dependent theory which can be combined  
with Gutzwiller-DFT computations, the concept 
of the TDGA has to be generalized to multiband Hubbard models 
with arbitrary local interactions. 
This is the main purpose of the present paper.
 
We consider a general class of multi-band Hubbard models which is    
defined as                                                                     
\begin{equation}\label{1.1}                                                    
\hat{H}=\sum_{i \ne j;\sigma,\sigma'}t_{i,j}^{\sigma,\sigma'}                  
\hat{c}_{i,\sigma}^{\dagger}\hat{c}_{j,\sigma'}^{\phantom{+}}                  
+\sum_i \hat{H}_{{\rm loc},i}=\hat{H}_0+\hat{H}_{\rm loc}\;.                   
\end{equation}                                                                 
 Here, the first term describes the hopping of electrons between $N$           
spin-orbital                                                                   
states $\sigma,\sigma'$ on~$L$ lattice sites $i,j$, respectively. The Hamiltonian  $\hat{H}_{{\rm loc},i}$
contains all local terms, i.e., two-particle Coulomb interactions  
and orbital onsite-energies. 
We further introduce the eigenstates $|\Gamma_i \rangle$ of  
$\hat{H}_{{\rm loc},i}$ and the corresponding energies $E^{\rm loc}_{\Gamma,i} $, i.e. 
$\hat{H}_{{\rm loc},i} |\Gamma_i \rangle = E^{\rm loc}_{\Gamma,i}  |\Gamma_i \rangle 
\;$.

Within the Gutzwiller theory~\cite{gutz,buen98},  
the Hamiltonian~(\ref{1.1}) is investigated by  
means of the variational wave function                                        
$|\Psi_{\rm G}\rangle=\hat{P}_{\rm G}|\Psi_0\rangle 
=\prod_{i}\hat{P}_{i}|\Psi_0\rangle\;$, 
where $|\Psi_0\rangle$ is a normalized single-particle product state and the 
local Gutzwiller correlator 
$\hat{P}_{i}=\sum_{\Gamma,\Gamma^{\prime}}\lambda^{(i)}_{\Gamma,\Gamma^{\prime}} 
|\Gamma \rangle_{i} {}_{i}\langle \Gamma^{\prime} |$ 
depends on the matrix $\tilde{\lambda}$ of variational parameters  
$\lambda_{\Gamma,\Gamma^{\prime}}$.

The expectation value of the Hamiltonian (\ref{1.1}) in infinite dimensions 
(i.e., in GA) is read as 
 \begin{eqnarray} 
E^{\rm GA}(\tilde{\rho},\tilde{\lambda}) &=&  
\sum_{i \ne j;\sigma,\sigma',\sigma'_1,\sigma'_2} 
t^{\sigma'_1,\sigma'_2}_{i,j} q_{\sigma'_1}^{\sigma} 
\left( q_{\sigma'_2}^{\sigma'}\right)^{*} \rho_{j\sigma',i\sigma} 
\nonumber \\\label{1.4b} 
&+& L\sum_{\Gamma,\tilde{\Gamma},\tilde{\Gamma}'}              
E^{\rm loc}_{\Gamma}\lambda^{*}_{\tilde{\Gamma},\Gamma} 
\lambda_{\Gamma,\tilde{\Gamma}'}^{}m^0_{\tilde{\Gamma},\tilde{\Gamma}'} \label{1.13a} 
\end{eqnarray} 
where $m^0_{\Gamma,\Gamma'}\equiv\langle \left( |\Gamma \rangle   \langle   
\Gamma' |\right)  \rangle_{\Psi_{0}}$ can be expressed~\cite{buen98} via the 
local part of the uncorrelated density matrix $\tilde{\rho}$ with the elements 
$\rho_{j\sigma',i\sigma}\equiv\langle \hat{c}_{i,\sigma}^{\dagger}\hat{c}_{j,\sigma'}^{\phantom{+}}\rangle_{\Psi_0}$. The `renormalization factors' $q_{\sigma'}^{\sigma}$ in (\ref{1.4b}) 
account for the correlation induced band narrowing and depend on 
$\tilde{\lambda}$ and $\tilde{\rho}$.

The energy functional Eq.\ (\ref{1.13a}) has to be minimized with respect 
to the $n_{\rm var}$ variational parameters $\lambda_{\Gamma,\Gamma'}$  
and the elements of the density matrix $\tilde{\rho}$~\cite{slater}. 
Within the framework of the GA the variational parameters 
 have to obey 
the completeness condition  
$C_i^{(1)}(\tilde{\rho},\tilde{\lambda}) \equiv \langle\hat{P_i}^{\dagger}\hat{P_i}^{}\rangle_{\Psi_0} - 1 = 0\;$ plus $n_{\rm con}$  further local 
constraints      
$C_{i,\sigma\sigma'}^{(2)}(\tilde{\rho},\tilde{\lambda}) \equiv   
\rho_{i\sigma',i\sigma} -  
\langle \hat{P_i}^{\dagger} \hat{P_i}^{} \hat{c}^{\dagger}_{\sigma}                
\hat{c}^{}_{\sigma'} \rangle_{\Psi_0} = 0$ 
which suggests the definition of the Lagrange function 
\begin{eqnarray} 
{\cal L}^{\rm GA}(\tilde{\rho},\tilde{\lambda}) &=&  
E^{\rm GA}(\tilde{\rho},\tilde{\lambda}) 
+\sum_i \Lambda^{(1)}_i C_i^{(1)}(\tilde{\rho},\tilde{\lambda}) \nonumber \\ 
&+&\sum_{i,\sigma\sigma'}\Lambda^{(2)}_{i,\sigma\sigma'} 
C_{i,\sigma\sigma'}^{(2)}(\tilde{\rho},\tilde{\lambda}). \label{eq:lag} 
\end{eqnarray} 
The GA ground state is thus defined from the condition $\delta {\cal L}=0$ 
which fixes all $n_{\rm var}$ variational parameters  
$\lambda_{\Gamma,\Gamma'}$ plus the $n_{\rm con}+1$ Lagrange parameters  
$\tilde{\Lambda}=\{\Lambda^{(1)}_i ,\Lambda^{(2)}_{i,\sigma\sigma'}\}$.    
Having thus determined the GA ground state, one can 
define an auxiliary `Gutzwiller single-particle  
Hamiltonian' 
\begin{equation}\label{hga} 
H_{i\sigma,j\sigma'}^{\rm GA} 
=\frac{\partial {\cal L}^{\rm GA}}{\partial \rho_{j\sigma',i\sigma}} 
\end{equation} 
the diagonalization of which yields also the unoccupied single-particle 
orbitals in addition to the occupied ones already implicit in 
the ground-state density matrix $\tilde{\rho}^0$.

A small time-dependent perturbation will induce small (time-dependent) 
deviations in both the density matrix ($\delta\tilde{\rho}$) and   
in the variational and Lagrange parameters  
($\delta\tilde{\lambda}$, $\delta\tilde{\Lambda}$) from their ground state 
values. Following the spirit of time-dependent HF theory (TDHF) 
[see e.g. 
\cite{RING,BLAIZOT}] the response of the system can be obtained from an  
expansion of the energy functional up to second order in the density  
fluctuations. The complication in the present case is due to the additional 
fluctuations in the variational parameters whose time-dependence is unknown. 
In case of the one-band model one can use 
the constraints to express all variational parameter fluctuations 
but one 
(which is usually 
chosen as the double occupancy fluctuation $\delta d$) 
via the density fluctuations $\delta\rho$
(and $\delta d$).  
In the multi-band case studied here, one usually has  
$n_{\rm var} \gg n_{\rm con}$ and there is no general analytical way  
to express some $n_{\rm con}+1$ dependent variational parameters via 
the remaining $n_{\rm var}-n_{\rm con}-1$ independent ones and the density matrix.  
This problem can be circumvented by the fact  
that the expansion of $E^{\rm GA}$ in 
the independent parameter fluctuations is equivalent to an 
expansion of the Lagrange function Eq.\ (\ref{eq:lag}) in {\it all}  
the fluctuations $\delta\tilde{\rho}$ and $\delta\tilde{\lambda}$. 
Expanding Eq.\ (\ref{eq:lag}) 
up to second order yields 
\begin{eqnarray} 
\delta{\cal L}^{\rm GA}&=&{\rm Tr}(\tilde{h}^0 \delta\tilde{\rho})  
+ \frac{1}{2} \frac{\partial^2{\cal L}^{\rm GA}} 
{\partial\rho_i\partial\rho_j}\delta\rho_i\delta\rho_j \label{no6} \\ 
&+&\frac{\partial^2{\cal L}^{\rm GA}} 
{\partial\rho_i\partial\lambda_j}\delta\rho_i\delta\lambda_j
+ \frac{1}{2} \frac{\partial^2{\cal L}^{\rm GA}} 
{\partial\lambda_i\partial\lambda_j}\delta\lambda_i\delta\lambda_j \nonumber \\
&+&\sum_n \delta\Lambda^{(1)}_n\left\lbrace  
\frac{\partial C_n^{(1)}}{\partial\rho_i}\delta\rho_i 
+ \frac{\partial C_n^{(1)}}{\partial\lambda_i}\delta\lambda_i\right\rbrace  \nonumber \\
&+& \sum_{i,\sigma\sigma'}\delta\Lambda^{(2)}_{n,\sigma\sigma'}\left\lbrace 
\frac{\partial C_{n,\sigma\sigma'}^{(2)}} 
{\partial\rho_i}\delta\rho_i 
+ \frac{\partial C_{n,\sigma\sigma'}^{(2)}} 
{\partial\lambda_i}\delta\lambda_i\right\rbrace \nonumber. 
\end{eqnarray}
where  $\tilde{h}^0$ is given by the GA ground-state result for 
(\ref{hga}).
For clarity, we have casted the multiplicity 
of dependencies of the fluctuations in a single index. 
Like in the single-band case~\cite{goe01}
we can now relate the dynamics 
of the  $n_{\rm var}+n_{\rm con}+1$ parameter 
fluctuations $\delta\tilde{\lambda}$ and $\delta\tilde{\Lambda}$ to that of 
the  
density fluctuations $\delta \tilde{\rho}$ 
requiring that  
$\frac{\partial{\cal L}^{\rm GA}}{\partial \delta\tilde{\lambda}}=0$
and  
$\frac{\partial{\cal L}^{\rm GA}} 
{\partial \delta\tilde{\Lambda}}=0$. 
The resulting linear system of equations can be inverted yielding  
$\delta\tilde{\lambda} 
=\delta\tilde{\lambda}[\delta\tilde{\rho}]$  
and $\delta\tilde{\Lambda} 
=\delta\tilde{\Lambda}[\delta\tilde{\rho}]$ which upon inserting into  
Eq.\ (\ref{eq:lag}) 
results in a Lagrange function which depends on density 
matrix fluctuations only 
$\delta{\cal L}^{\rm GA}(\delta\tilde{\rho})={\rm Tr}(\tilde{h}^0 
\delta\tilde{\rho})  +\frac{1}{2} 
{\cal K}_{ij} \delta\rho_i\delta\rho_j$ . 
A more general approach 
which explicitely incorporates the dynamics of the variational parameters
may be derived along the lines presented in \cite{fabrizio}. 
We expect, however, that 
this improvement will change the results mostly at high energies.    
 
The remaining steps are analogous to TDHF~\cite{RING,BLAIZOT}  
and TDGA for the one-band model~\cite{goe01}. We construct 
the `bare' susceptibilities $\chi^0_{ij}(\omega)$ 
on the GA level from the eigenstates 
of the Hamiltonian $\tilde{h}^0$. The corresponding 
correlation functions of the interacting system are then obtained 
from the RPA-like series $\chi_{ij}(\omega)=\chi^0_{ij}(\omega) 
+ \chi^0_{in}(\omega){\cal K}_{nm}\chi_{mj}(\omega)$. 
Note that , in contrast to conventional TDHF theory 
the TDGA induces also couplings 
between transitive density fluctuations  
(i.e., $\sim \delta\langle c_{i\sigma}^\dagger c_{j\sigma'}\rangle$)
which 
enlarges the size of the RPA-like matrix equation. 
Nevertheless,  
the numerical efforts of our method exceed those of an ordinary HF+RPA  
only marginally.   
 It is also straightforward to implement our approach into LDA+Gutzwiller 
schemes outlined in Refs.~\cite{ho08,deng08,deng09}.  The number of parameters 
 $\lambda_i$ needed in the expansion (\ref{no6}) depends on the symmetries of 
the model and the response function one aims to calculate. In our calculations
 of magnetic susceptibilities for a two-band model (see below) 
 $16$ such parameters have been used. 
 
We exemplify our method for a Hamiltonian with two degenerate 
bands on a hypercubic (hc) lattice (intra-orbital  
hopping $t_{ij}=t/\sqrt{2{\cal D}}$) where the GA is exact for the  
 Gutzwiller variational wave-function. Note that for the hc lattice  
the momentum dependence of the susceptibilities is contained in a  
single parameter 
$\eta=\eta_\qvec=\frac{1}{\cal D} [\cos(q_1)+\cos(q_2)+ ....+\cos(q_D)]$  
\cite{MH} which takes values $-1 \le \eta \le +1$. 
We work with the standard atomic Hamiltonian for two degenerate $e_g$ orbitals 
~\cite{buen98} with intra-orbital Coulomb interaction $U$  
 and exchange interaction $J$.  
Note that 
in contrast to most previous DMFT investigations of similar 
models (cf. Ref.~\cite{peters10} and references therein)    
we explicitly include 
the pair hopping term  
to guarantee rotational invariance in 
orbital-spin space.  
\begin{figure}[ttt] 
\includegraphics[width=8.5cm,clip=true]{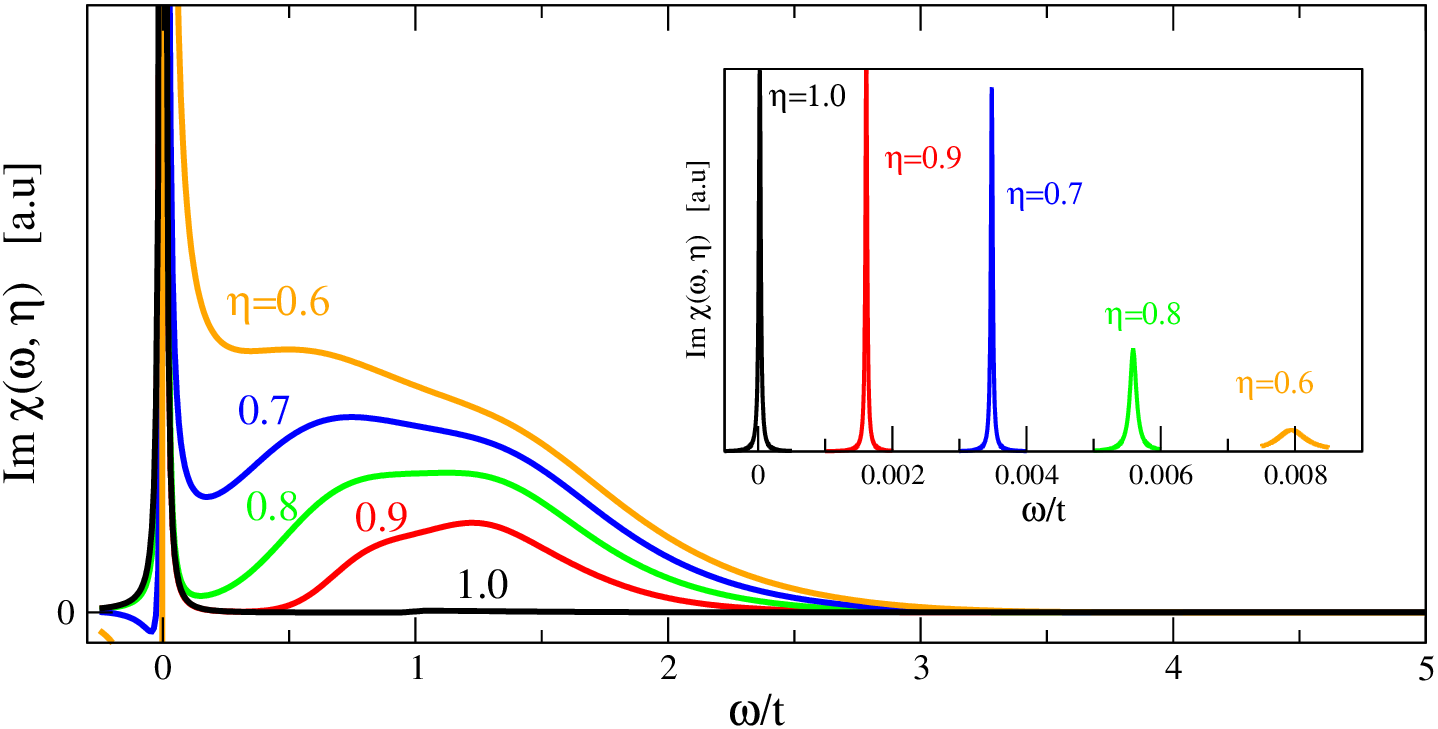} 
\caption{(color online) Frequency and momentum dependence of the magnetic susceptibility 
for finite ferromagnetic ground-state magnetization $n_\uparrow-n_\downarrow=0.677$ per orbital, various values of $\eta$, 
doping $\delta=0.6$, and $U/t=10.5$, $J/U=0.2$. The main panel covers 
the frequency range up to the Stoner excitations whereas the inset 
shows the low energy magnon part. 
} 
\label{fig3} 
\end{figure} 
Fig.\ \ref{fig3} displays the  
spin susceptibility 
$\chi(\qvec,\omega)=\int dt exp(i\omega t)\langle {\cal T} 
\hat{S}^{+}_{\qvec}(t) \hat{S}^-_{\qvec}(0)\rangle$ 
for a ferromagnetic system (magnetization $m=0.677$ per orbital) 
where $S^+_{\qvec}$ denotes the spin-flip operator in momentum space. 
The spectrum is composed of a low energy magnon (cf. inset to Fig.\ \ref{fig3})  
and the higher energy Stoner contribution.  
In the limit $\qvec\to 0$ ($\eta \to 1$) and $\omega\to 0$   
one obtains the Goldstone mode as a delta-function (broadened by  
$\epsilon=10^{-5}t$) in Fig.\ \ref{fig3}) which comprises all the spectral  
weight and which 
provides an additional consistency check of our approach.  
For finite momenta the magnon excitations get rapidly damped (cf. inset to 
Fig.\ \ref{fig3}) due to the large phase space of spin flip particle-hole 
decay processes in infinite dimensions. 
Contrary to HF+RPA, where the Stoner excitations are expected to  
appear at $\omega \sim (U+J) m $ ($=8.53t$ for the  
same magnetization) one observes that these are reduced to $\sim 1.22t$ 
within the TDGA thus rapidly merging with the magnon excitations for 
$\eta <1$. With regard to this correlation induced reduction of the 
Stoner parameter, our findings are similar to local 
Ansatz investigations~\cite{oles91} but they generalise them by providing  
us with the full spin-excitation spectrum. 
 
\begin{figure}[ttt] 
\includegraphics[width=8.5cm,clip=true]{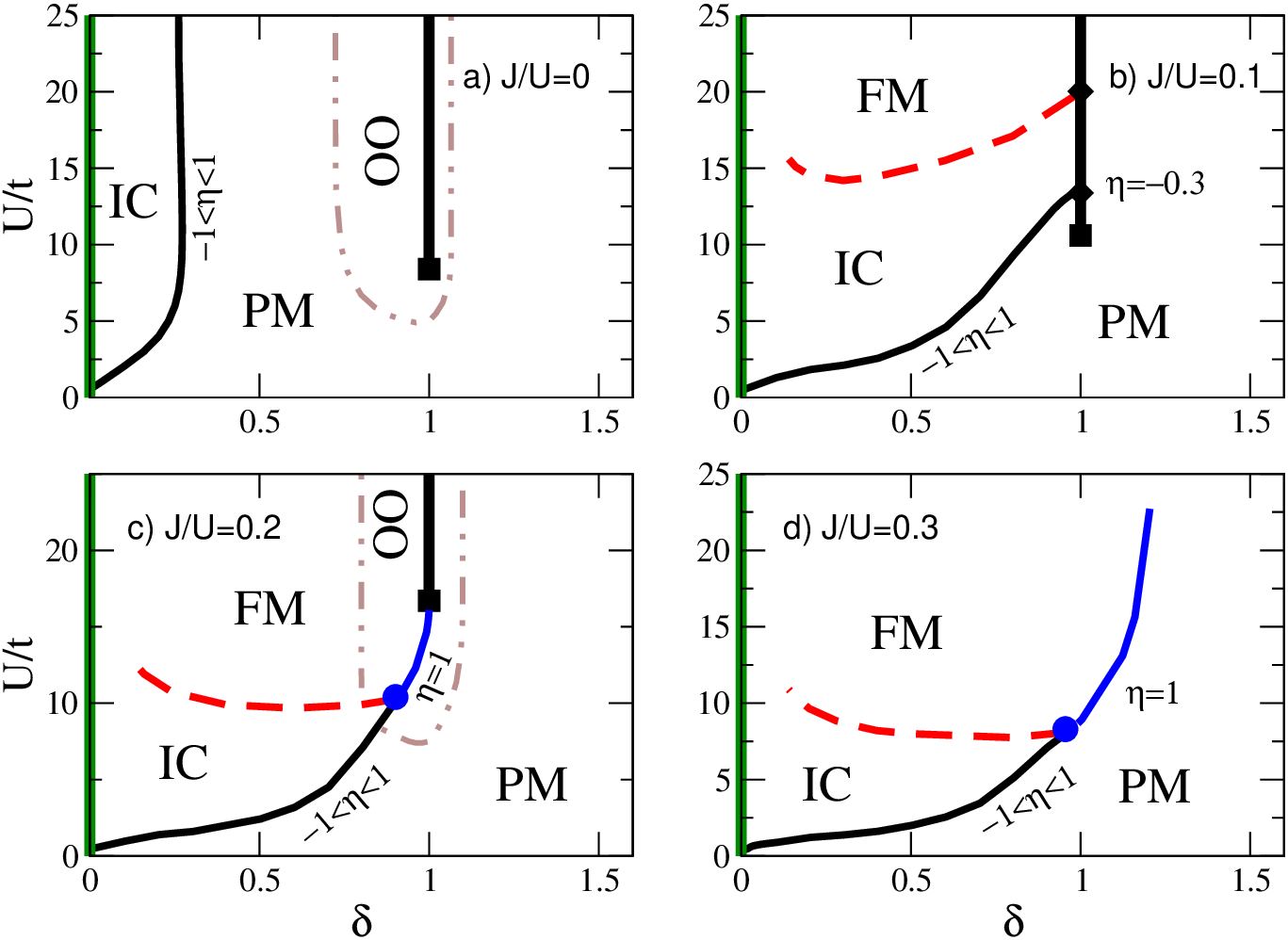} 
\caption{(color online) TDGA magnetic phase diagram for the two-band model on  
the hypercubic lattice. Shown are the phase boundaries between paramagnetic  
(PM), incommensurate (IC) and stable ferromagnetic (FM) phases.  
The full square at $\delta=1$ indicates the BR transition  
towards a localized regime (thick solid line). 
The solid circle marks a 
Lifshitz point where PM, IC, and FM phases merge.  
AF order is stable at $\delta=0$ (solid green line).  
In panels (a,c) also the 
orbital ordered (OO) phase is shown (dashed dotted). 
Doping $\delta$ is defined 
as the concentration of holes per site with $\delta=0(1)$ being the 
state with $2(1)$ particles per site.} 
\label{fig1} 
\end{figure} 
 
\begin{figure}[bbb] 
\includegraphics[width=8.5cm,clip=true]{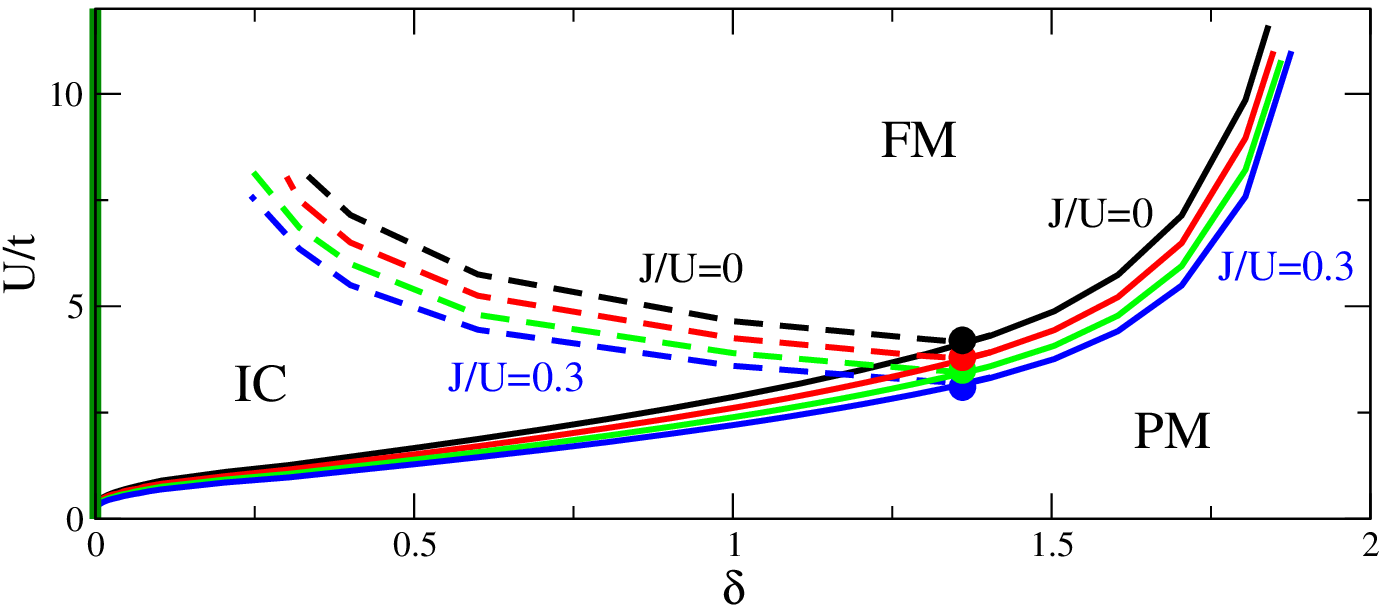} 
\caption{Same as Fig.\ \ref{fig1} but obtained within HF+RPA.} 
\label{fig2} 
\end{figure}

Figs.~\ref{fig1} and \ref{fig2} report the magnetic phase diagrams  
for both TDGA and HF+RPA and various values of $J/U$, respectively.  
We have determined the second order 
transition between paramagnetic (PM) and incommensurate (IC) or PM and 
ferromagnetic (FM) phases from the 
condition ${\rm DET}|1 - \chi^0_{in}(\omega=0){\cal K}_{nm}|=0$.
 For the PM-FM transition we verified the second-order nature by 
calculating the ground-state energy as a function of the magnetic 
moment. 
Especially the evaluation of IC phase boundaries is straightforward 
in our approach whereas the local nature of DMFT has hampered the 
determination of these phases in related investigations  
\cite{held98,momoi98,chan09}.  
The solid lines in Figs.~\ref{fig1} and \ref{fig2} 
are the enveloping curve for all the phase 
boundaries with different $\eta_\qvec$. At half filling ($\delta=0$) 
 an antiferromagnetic 
(AF) phase  is stable over the whole range of $U/t$ ($\eta=-1$)  
 due to perfect nesting. 
 It evolves 
continuously into IC ($-1<\eta<1$) and eventually FM ($\eta=1$)  
 phases at finite doping. 
Stability of the latter has been determined from the magnetic excitation 
spectrum (cf. Fig.\ \ref{fig3}). The dashed lines in Figs.~\ref{fig1} and  
\ref{fig2} separate a stable FM state with positive `spin-wave stiffness'  
(i.e. a positive slope of the magnon dispersion in the ferromagnetic state) 
from the IC phases where a FM state would decay due to quantum fluctuations. 
Note that due to the stability of the AF phase at $\delta=0$  
these phase boundaries  
diverge as a function of $U/t$ upon approaching half-filling.  

The HF+RPA phase diagram shows the known limitations of a mean-field 
multiband approach: It depends  only weakly on $J/t$ and it contains 
a stable FM state in the whole doping regime at sufficiently large $U/t$. 
In addition, one always finds a a Quantum Lifshitz 
point (QLP) where IC, PM, and FM phases meet, similar to previous 
investigations in the one-band Hubbard model on the hc lattice 
\cite{falk07}. In contrast, within the TDGA
 a stable FM phase requires a minimum value of $J/U$ (cf. Fig.\ \ref{fig1}a)  
which favors local triplet formation 
in agreement with DMFT~\cite{held98}. 
In the intermediate $J/U$ regime (Figs.~\ref{fig1}b,c) IC and FM phases 
are bounded by the Brinkman-Rice (BR) localization transition which occurs 
at $\delta=1$ and $U_c/t$ (the latter depending on $J/U$). Note that  
the same termination of the FM phase has been  
found in DMFT investigations on a three-orbital model~\cite{chan09}.  
Mapping to a Kugel-Khomskii Hamiltonian \cite{kugel73} suggests the 
formation of additional orbital order (OO). This OO is expected     
around  doping  $\delta=1$ being FM in the spin- and                
AF in the orbital channel, i.e.                                     
$(\circ\uparrow),(\uparrow\circ),(\circ\uparrow),(\uparrow\circ)$.  
DMFT investigations of this issue within the                        
two-orbital model remain ambiguous \cite{held98,momoi98}. In         
Fig. \ref{fig1}a,c the OO phase as obtained from minimizing the GA functional
is shown which 
in the PM regime occurs as a first order transition. Within the FM  
phase the transition towards OO is second order for $J/U=0.2$. However, 
we find that it can also be of first order for larger values $J/U=0.3$  
(not shown). For all values of $J/U$ the BR transition 
at $\delta=1$ is masked by OO. It thus appears as a generic  
finding that PM insulating phases at integer doping are prevented by another 
instability, either magnetic (AF at $\delta=0$) or due to OO  
(at $\delta=1$), similar to the three-orbital model~\cite{chan09}. 
 
The multiband TDGA  can be easily generalized 
to the pair and charge sector where in the latter it allows for the  
computation of response functions like optical                       
conductivity and polarizability for multiorbital                     
Hubbard models.                                                      
It is obvious                                                   
that the formalism offers a valuable tool with manageable numerical effort 
for the investigation 
of the dynamics of multiband Hubbard models in the context of correlated 
materials. 
 
We acknowledge financial support from the Deutsche Forschungsgemeinschaft.


\begin{thebibliography}{10}
\bibitem{slater} The variations of $\tilde{\rho}$  are subject to the  
  constraint that it is constructed from a Slater determinant.
\end{thebibliography}
\end{document}